# Ultra-slow sound in non-resonant meta-aerogel

Ai Du[1,†,*], Yuhan Xie[1,†], Yan Li[2], Yongdong Pan[2], Guangming Wu[1], Bin Zhou[1]

**Abstract**

The manipulation of sound with acoustic metamaterials is a field of intense research, where interaction via resonance is a common application despite the significant disadvantages. We propose a novel procedure for introducing well-designed coupling interfaces with a cell size of less than 10 nm into an ultra-soft porous medium, to prepare a meta-aerogel, where the sound propagation is significantly delayed in a non-resonant mode. The resultant sound velocity is shown as a scaling law with the mass density and the mass fraction ratio of the components, in accordance with our analytical model. We have prepared a meta-aerogel with the slowest sound velocity of 62 m/s. To the best of our knowledge, this is the lowest value in compact solid materials, with a prospect of further slowing down by our procedure. The development of such meta-aerogels can facilitate key applications in acoustic metamaterials intended to employ non-resonant type slow sound (or phase delay). Examples of the latter include deep subwavelength meta-surface and other focused imaging or transformation acoustics that require a high contrast of sound velocity.

[1] Shanghai Key Laboratory of Special Artificial Microstructure Materials and Technology, School of Physics Science and Engineering, Tongji University, Shanghai 200092, China
[2] School of Aerospace Engineering and Applied Mechanics, Tongji University, 200092 Shanghai, China
† These authors contributed equally: Ai Du, Yuhan Xie.
* Correspondence and requests for materials should be addressed to A. D. (email: duai@tongji.edu.cn).

**Introduction**

Slowing the sound propagation speed has become the cutting edge work in the field of designing the acoustic metamaterials. Since the first experimental configuration of locally resonant sonic material by Liu et al.[1], subsequent works on acoustic metamaterials focused on the deep subwavelength acoustic single or double negativity[2] by resonant interaction. The disadvantages of a narrow band and low efficiency at low frequency regions are significant. This can be alleviated with multi-layered composites of a slow sound medium[3]. Considering the physics of the acoustic double negativity[4], proposals for introducing Mie resonance[5] and coiling up space[6] are currently the most common. For the coiling-up method, the slow sound is a demanding characteristic, while in Mie resonance the high contrast of sound velocity is the crucial dominator. However, no homogeneous natural material possesses a sufficiently low sound speed to exhibit such features. Investigations of slow sound have been conducted in wave-guided sonic crystals[8], by considering sound propagation in pipes with a series of detuned Helmholtz resonators[7] and in dissipative resonators[9], all of which are due to the narrow-band resonance with an inevitably large size.

Nanomaterials can exhibit strong interaction with the incident wave, making a large slow down[10,11] in the propagation. Sheng has investigated in detail the wave propagation in the mesoscopic strong scattering medium[12,13], indicating that the propagating wave will quickly change into diffusive transport, whose wave front is dismissed, and hence they primarily propagate via scattering process. Hence it is noticeable that by introducing a resonance to enhance the interaction with the microstructure, the wave is slowed down or even localized. However, the diffusive transport is unnecessarily a perpetually resonant type. Several works have achieved slow sound in water-dispersing glass beads[12], granular medium (e.g., sand)[14], and porous media[15]. Even the lowest sound speed about 100 m/s has been achieved in porous silica rubber[16].

Ultra-soft nanomaterial, such as aerogels, would be a good breaker. A series of studies have found[17] that the aerogels exhibit a scaling law between the sound velocity and mass density. The lowest sound velocity of about 100 m/s has been observed in silica aerogels with the densities of about 20–50 kg/m$^3$. However, the sound velocity increases with the decrease of the mass density when it is below 20 kg/m$^3$. The pore size increases so that the propagation via the air dominates accordingly. Similarly, owing to the ultra-soft property attached with the ultralow shear modulus, soft silica rubber with less porosity can possess a lower sound speed downwards 80 m/s. But pursuing the goal of further slowing down reinstates the importance of studies on high porosity[16].

Here, we propose a feasible solution for reinforcing the non-resonant type interaction that depends on two factors, viz., (1) nanosized scattering units, which produce high filling fraction and large amounts of coupling interfaces, and (2) the well-designed mechanical properties of a single unit. We introduce highly dense coupling interfaces, for a cell size smaller than 10 nm, into the ultra-soft aerogel to attain a meta-aerogel with ultra-slow sound propagation. By constructing a well-designed microstructure, the sound propagation can be significantly slowed down, resulting in the lowest velocity ever (to our knowledge) in compact solids at 62 m/s. Notably, owing to the fact that the ultrasonic incident wave is delayed in a nanoscale microstructure, the slow sound is no longer confined to a resonant-type phenomenon, thus paving the way for non-resonant slow sound and key applications of wide-band and efficient metamaterials.

**Results**

*Observation of low-sound-velocity*

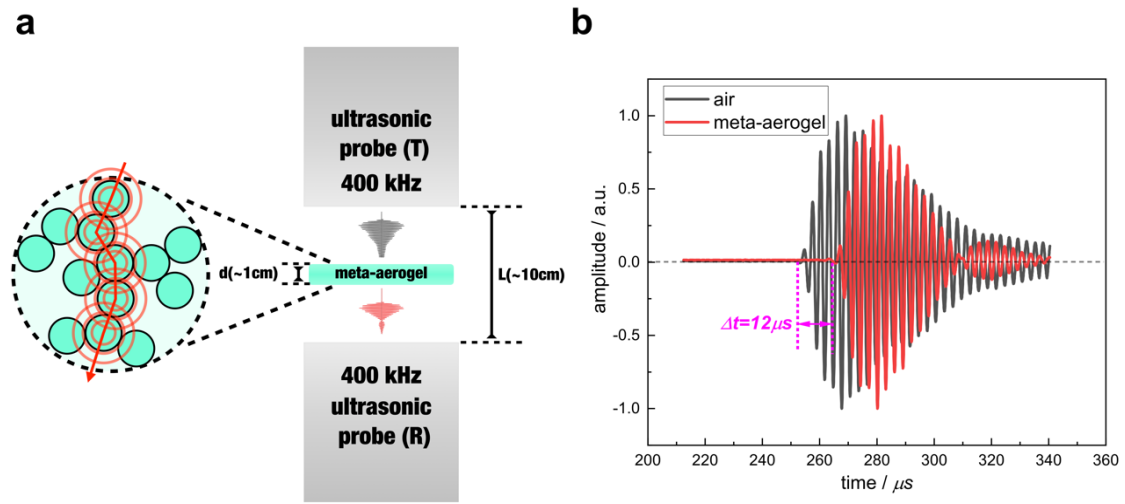

*Figure 1 | Observation of the low-sound-velocity. **a**, Illustration of sound velocity testing installation, consisting of two focused non-contact air-coupled ultrasonic probes, whose working frequency is 400 kHz (photos are shown in SI. A). **b**, Example of measured ballistic coherent pulses transmitted through the meta-aerogel.*

Observed from the air-coupled ultrasound experiment (Fig. 1), the sound is significantly delayed while traveling through the meta-aerogel. It can be derived from the transition equation (SI. A) that the sound velocity in the meta-aerogel is slower than that in the air. The sound velocity is 209 m/s in this sample with a density of 340 kg/m$^3$ (Fig. 1b). However, according to the scaling-law of conventional silica aerogel, the sound velocity with that mass density should be around 500 m/s[17], which implies that we have preliminarily achieved our goal, i.e., artificially slowing the sound. Indeed, the sound propagation in the sample that we have tested here is no longer a coherent sound wave but a diffusive transport (Fig. 1a). The transport process will be discussed in detail in our analytical model.

*Configuration of the meta-aerogel with ultra-slow sound velocity*

The basic idea to enhance the interaction between the sound wave and the material in a non-resonant way is to introduce nanoscale heavy scatterers into the ultra-soft substrate in the aerogel. This is done by adjusting the phase-separation mode of the copper(II)/polyacrylic acid (Cu(II)/PAA) hybrid systems. In our previous studies[18,19], PAA could confine the nucleation sites of inorganic compounds by coordinating with the metal ions, and hence facilitate the formation of the hierarchical gel structure via dispersed inorganic sol-gel method. Then the meta-aerogels have been prepared via the combination of pinhole ambient drying. Further, the ambient drying method at 45 °C could produce the aerogels with relatively high and similar densities since the gels have similarly high shrinkage ratios. Concomitantly, we have applied the supercritical drying method, which can efficiently maintain the microstructure of the gel and obtain a low-density aerogel. It is statistically shown in Fig. 2a that the meta-aerogel consists of a typical 2-level hierarchical microstructure. The primary structure with a size of about 8 nm consists of a hard core embedded in a soft shell of PAA. The heavy core scatterer inside the PAA is the primary particle with a statistical average size of 2.5 nm, composed of copper chloride hydroxide ($Cu_2(OH)_3Cl$). This primary structure can be seen as a typical core-shell structure and notably its size is only about several nanometers.

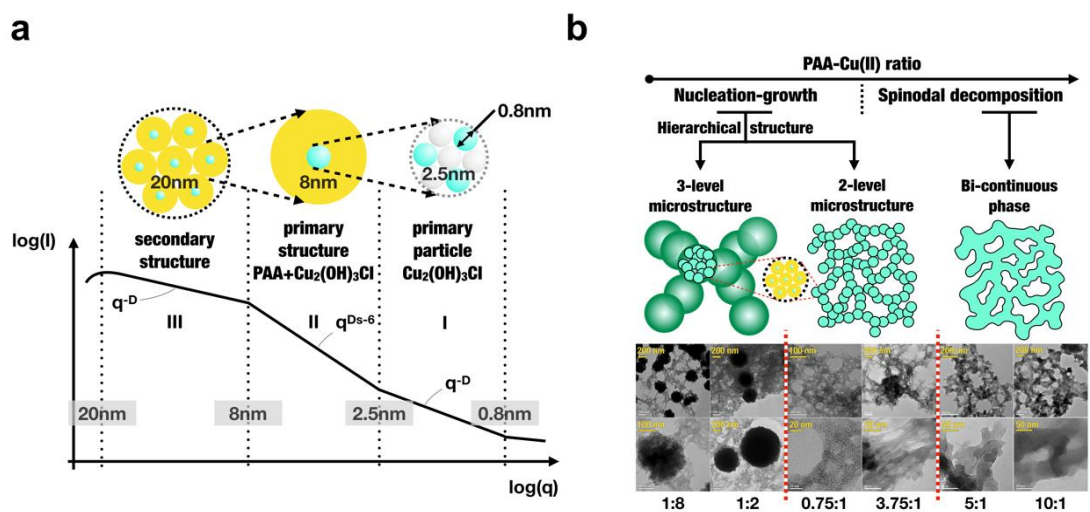

*Figure 2 | Typical hierarchical microstructure of meta-aerogel. a, Typical SAXS graph of the meta-aerogel, statistically showing the hierarchical microstructure of different fractal dimensions. b, Three periods of the microstructure of the meta-aerogel with an increasing amount of PAA. "PAA-Cu(II) ratio" accounts for the mole ratio of PAA monomer to Cu ion.*

By adding different amounts of PAA, we can continuously adjust the hierarchical microstructure of the aerogel. Specifically, when adding a small amount of PAA, a higher level of self-assembly occurs in the aging process of Cu(II)/PAA systems (microsyneresis), leading to a three-level hierarchical microstructure. When the mole ratio of PAA monomer to Cu ion (denoted as "PAA-Cu(II) ratio" hereafter) is in the range of 0.125~0.5, as shown in Fig. 2b, submicron sphere skeleton (tertiary structure) is formed via sol-gel transition and dehydration-induced self-assembly during the aging process. The secondary structure of the size of several tens of nanometers could be verified in the zoomed-in TEM graphs of the gels (Fig. 2b). However, for the samples with a PAA-Cu(II) ratio of 0.75 and 3.75, an obvious two-level structure could be observed. Further amounts of PAA could react with the surface hydroxyl groups and stop the interparticle-dehydration-induced self-assembly, resulting in a two-level structure. Also, a larger amount of PAA (PAA-Cu(II) ratio no less than 5) has been added to form a bi-continuous porous structure via spinodal decomposition, taking a cue from the works of Nakanishi's group[20]. As shown in the Fig. 2b, there is no fine structure of the skeletons.

Therefore, we can continuously adjust the microstructure of the meta-aerogels from three-level and two-level hierarchies to a bi-continuous structure by changing the amount of added PAA. Furthermore, as shown in SI. B, we can also adjust the mass or surface fractal dimensions and the average size of the primary and secondary microstructure (Fig. S2) by designing the mass density and PAA-Cu(II) ratio, to get the target properties of the meta-aerogels.

*Scaling law of velocity vs. density & mass fraction ratio of components*

As commonly seen in conventional aerogel, the meta-aerogel also possesses a scaling law between the sound velocity and mass density. By pinhole ambient drying, we carefully adjust the mass densities of the meta-aerogels with the same PAA-Cu(II) ratio, which contributes to a confined microstructure but a significantly altered concentration of interaction units. As shown in Fig 3a, the sound velocity is effectively slowed down with the decrease of mass density, following the scaling law. Excitingly, the speed value of the meta-aerogel can be as low as 62 m/s when its density and PAA-Cu(II) ratio are 160 kg/m$^3$ and 1, respectively.

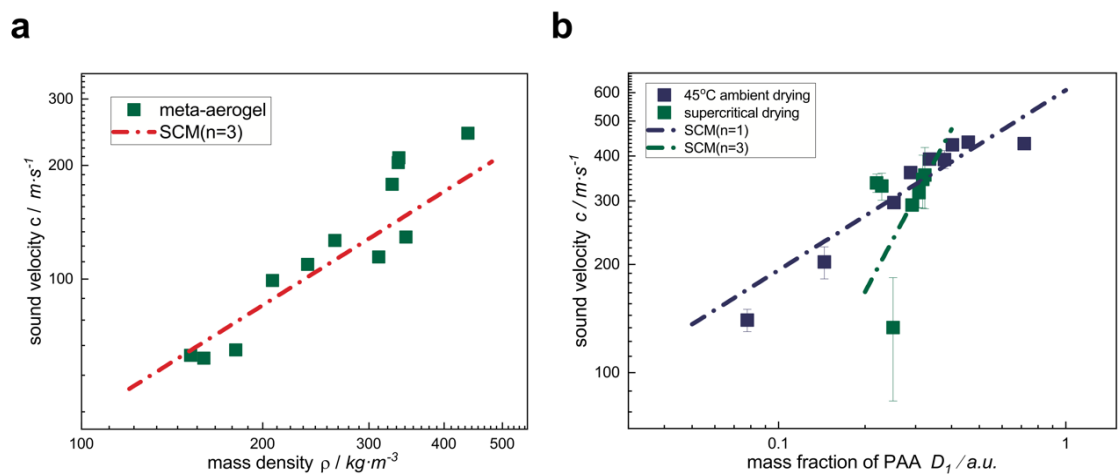

*Figure 3 | Scaling law of a, sound velocity vs. apparent mass density of meta-aerogel consisting of the same components; b, sound velocity vs. mass fraction of PAA. The theoretical prediction is also plotted in the graph as the strong-coupling-model (SCM). The predicted index (n) 1 and 3 for Fig. 3b correspond to 45 °C ambient drying and supercritical drying (SI. B), respectively.*

Apart from the apparent mass density, i.e., the PAA-Cu (II) ratio is a more important parameter to slow down the sound propagation since it directly influences the degrees of interaction of a single unit. Hence we change the mass fraction of PAA while maintaining the apparent density of aerogel, and then we prepare the aerogels by both the supercritical and 45 °C ambient drying methods. As predicted from Fig. 3b, the sound velocity of meta-aerogels shows a scaling law with the mass fraction of PAA. The experimental data is consistent with the theoretical prediction, which is described in detail below. Unconventionally, the resulting aerogels by ambient drying method can possess a relatively high density at about 1600 kg/m$^3$, whereas the sound velocity is low at nearly 100 m/s for a very low mass fraction of PAA. The difference made by the drying methods is related to the changes in the characteristics of the microstructure prepared (see Fig. S3) such as the fractal dimensions and the average size of the porous PAA and inner $Cu_2(OH)_3Cl$. Such a structural change also affects the indexes (SI. B).

*Theoretical model of sound propagation in meta-aerogel*

To understand further the ultra-slow sound propagation behaviors in our meta-aerogel, we construct theoretical models as illustrated in Fig. 4. Since the wavelength we use in the sound velocity test exceeds the size of the microstructure, the preliminary idea is to apply the effective medium method by considering the structure which carries the sound propagation as a two-component foam, i.e., porous PAA embedded by $Cu_2(OH)_3Cl$ core. Owing to the theory of wave propagation in mesoscopic medium[10], the incident wave would be soon transformed into diffusive transport after transmitting through several mean free paths. Hence, the object that we focus on here is the transport process via stress propagation along the arrangement of scatterers. We construct a concentrated-mass model, and use the electrical analogy to intuitively analyze its microscale dynamics.

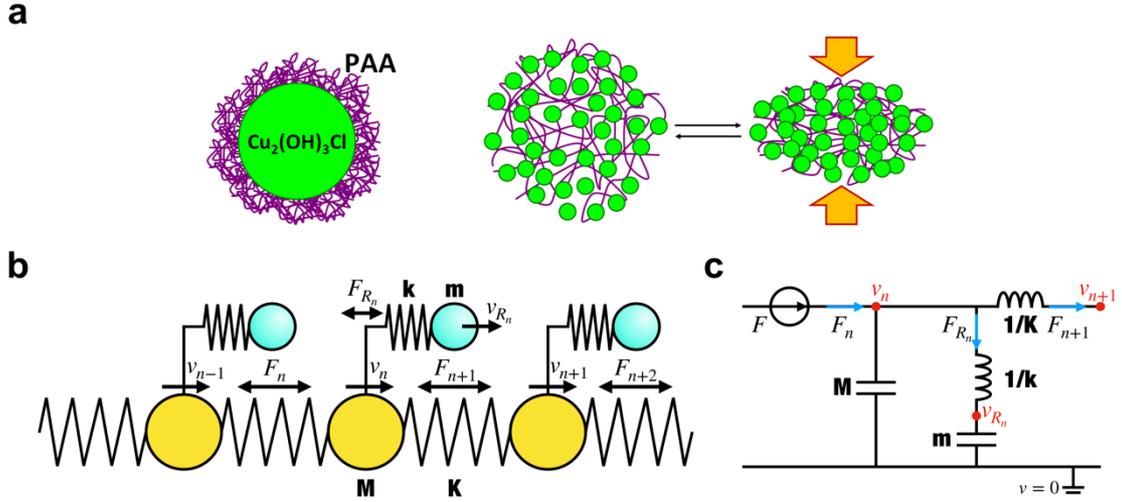

*Figure 4 | Physical model of meta-aerogel interacting with the sound wave. a, The illustration of the effective medium model of the meta-aerogel. b, Illustration of the analytical model of sound propagation in the microstructure of meta-aerogel. c, Illustration of the electrical analogy of the model.*

*(1) Effective medium theory of elastic continuum solids*

From the effective medium perspective, the results of the sound velocity can be explained from calculating the effective parameters. Outside the "heavy core" compound $Cu_2(OH)_3Cl$, the porous PAA as the "soft shell" behaves as a loose "foam" (Fig. 4a). When the sound wave enters the meta-aerogel, stress loaded on the medium is spatially-periodic, and the heavy core hardly changes its shape while the porous PAA is strained primarily, owing to its low modulus. This implies that the PAA chiefly facilitates the elasticity while the core mainly provides the density. Thus, the modulus of porous PAA can be expressed as a power law of its apparent mass density, according to Gibson[21],

$$E_a = E_{PAA} = E_s \cdot \left(\frac{\rho_{PAA}}{\rho_s}\right)^n \qquad (1)$$

where $E_a$ is the apparent modulus of the aerogel, $E_{PAA}$ the modulus of porous PAA, $E_s$ the modulus of the pure dense PAA, $\rho_{PAA}$ the apparent density of porous PAA, and $\rho_s$ the density of the pure dense PAA. As for porous materials[18], the index $n$ is approximately between 1 and 3. Further, its apparent mass density can be shown to be the product of the mass fraction of PAA and the density of the aerogel as

$$\rho_{PAA} = D_1 \cdot \rho_a \tag{2}$$

where the $D_1$ is the mass fraction of PAA and $\rho_a$ is the density of the aerogel. Thus, the expression of the sound velocity $c_l$ is shown as

$$c_l = \sqrt{\frac{E_a}{\rho_a}} = \left(\frac{E_s}{\rho_s}\right)^{\frac{1}{2}} \cdot D_1^{\frac{n}{2}} \cdot \left(\frac{\rho_a}{\rho_s}\right)^{\frac{n-1}{2}} \tag{3}$$

where the velocity is mainly related to the apparent mass density of the aerogel and the mass fraction of PAA. Owing to the meta-aerogel mainly consisting of Cu(II) and PAA, which implies the sum of mass fraction of PAA ($D_1$) and $Cu_2(OH)_3Cl$ ($D_2$) being 1, the mass fraction of PAA can be replaced with the form of mass fraction ratio of "hard core" $Cu_2(OH)_3Cl$ and the "soft shell" PAA. Consequently, the expression can be rewritten as

$$c_l = \left(\frac{E_s}{\rho_s}\right)^{\frac{1}{2}} \cdot \left(1 + \frac{D_2}{D_1}\right)^{-\frac{n}{2}} \cdot \left(\frac{\rho_a}{\rho_s}\right)^{\frac{n-1}{2}} \tag{4}$$

*(2) Analytical model of the sound wave propagation in meta-aerogel*

To obtain the analytical model of the sound propagation in the meta-aerogel, we can simplify the propagation issue into the model of a one-dimensional vibration chain, assuming that the sound primarily propagates through the porous PAA (Fig. 4b). The modulus has been dominated by the outer soft shell PAA, whereas the inner hardcore is an extra component connected to the PAA, contributing to an extra interaction. The specific microstructure has been determined with the geometric parameters of both the shell and the hard core, and in this work the ratio of PAA and Cu(II) is mostly considered.

The sound velocity can be derived from the dispersion relation of vibration in this chain-like structure. The model is analyzed by considering the motion of the $n^{\text{th}}$ unit cell, which is governed by the equations of dynamics (SI. C). By including the force transported on the dynamical chain as the current transported in the circuit, as shown in Fig. 4c, we can get a more vivid view of the model for considering the interaction of the substrate and the attached unit. The aforesaid analogy (see SI.C) compares the spring and mass to an inductor and capacitor, respectively, and the stress in the chain is seen as the current whereas the velocity difference of the particles are correlated to the electric potential (Fig. 4c).

From the dispersion relation (SI. C) that follows from a few manipulations, we can obtain the sound velocity by dividing the incident wave angular frequency from the propagating wavenumber $q$ per unit length in this structure (SI. C) as

$$c = \frac{\omega d}{q} = \sqrt{\frac{E_{PAA}}{\rho_{PAA} d^2}} d \left(1 + \frac{\frac{m}{M}}{1 - \frac{\omega^2}{\omega_R^2}}\right)^{-\frac{1}{2}} \approx \sqrt{\frac{E_{PAA}}{\rho_{PAA}\left(1 + \frac{m}{M}\right)}} \quad (5)$$

where $m$ and $M$ are the mass of core and shell respectively, $d$ the length of a single unit, $q$ the wavenumber of the wave propagating in the aerogel, $\omega_R$ the angular Eigen frequency of the attached unit of about $5.2 \times 10^9$ rad/s (calculated in SI. C), and $\omega$ the angular frequency of the incident wave of about $2.5 \times 10^6$ rad/s. The results of such a model demonstrate that the slowness in sound propagation comes from the attached units, and it is closely related to the modulus ratio and the mass density ratio of each component inside a cell. Furthermore, the ratio of $\frac{\omega^2}{\omega_R^2} \approx 0$ indicates that a non-resonant solution is appropriate. Subsequent to applying linear approximation (SI. C), the expression becomes simpler and more concise, which intuitively shows that the sound velocity in the meta-aerogel is closely related to the mass ratio of the two major components, Cu(II) and PAA, which plays the essential role in enlarging the effective mass density of the structure, for the mass ratio $\frac{m}{M} > 1$.

Concomitantly, the microstructure of the meta-aerogel follows a strict hierarchical configuration as shown in the SAXS result (Fig. 2a), which implies that the mass ratio is roughly similar to the mass fraction ratio of the two components. Thus the expression can also be rewritten as

$$c = \sqrt{\frac{E_{PAA}}{\rho_{PAA}\left(1 + \frac{D_2}{D_1}\right)}} = \left(\frac{E_s}{\rho_s}\right)^{\frac{1}{2}} \cdot \left(1 + \frac{D_2}{D_1}\right)^{-\frac{n}{2}} \cdot \left(\frac{\rho_a}{\rho_s}\right)^{\frac{n-1}{2}} \qquad (6)$$

Therefore, both the analyses lead to an equivalent result. The reason for the equivalence is the proper linear approximation considered in the model, which implies a non-resonant interaction in the propagation (SI. C). Concurrently, the approximation according to the non-resonant interaction ($\frac{\omega^2}{\omega_R^2} \approx 0$) indicates the feasibility for treating the sound propagation via static mechanics ($\omega \to 0$) in the effective medium model.

The red dashed line in Fig. 3a shows the theoretical prediction from the analytical model, which is in good agreement with the experimental results. The error probably follows from the slightly variant configuration of the microstructure. According to Gibson's theory[18], the effective modulus of a porous medium primarily consists of hexagonal open pores behaves as $E_s(\frac{\rho_*}{\rho_s})^3$, and hence we use index 3. The theoretical result also indicates that an ultra-low Young's modulus by decreasing the apparent mass density of meta-aerogels can result in a low sound velocity. However, the blue and green dashed lines (Fig. 3b) show the well-corresponded theoretical prediction of relation between the sound velocity and the mass fraction of PAA, *n*=1 and 3, respectively, indicating that the aerogels produced by ambient drying method exhibits a relatively less ability to slow down the sound wave. The index of the blue dashed line is, probably, owing to the use of ambient drying method, and hence the open pores in the porous PAA inevitably becomes smaller and even closed (SI. B). Furthermore, the deviation in the data of supercritical drying method (Fig. 3b) shows that when the PAA-Cu(II) ratio is exceedingly low (<1), the microstructure will not maintain the status quo (Fig. 2b) but behaves as the meta-aerogel prepared from the ambient drying method. Based on the analytical model, it can be concluded that both the ultra-low Young's modulus of the base material and the high mass ratio of the components contribute to the ultra-slow sound propagation.

*Microscopic mechanism of ultra-slow sound velocity*

The analytical model that we have constructed demonstrates the relation between the sound velocity and the microstructure parameters of the meta-aerogel. However, as mentioned above, such a model also indicates that the ultra-slow sound propagation has no relation with the strong resonant-type interaction, since the angular frequency of the ultrasonic wave that we employ to test the sound velocity is only $2.5 \times 10^6$ rad/s. This angular frequency is far below that of the resonant region (see SI. C) of the microstructure cell that we introduce, which is about $5.2 \times 10^9$ rad/s. Apparently, the phenomenon of ultra-slow sound should be the result of a type of interaction exclusive of the resonance. Notably, the mass fraction ratio of Cu(II) and PAA plays an important role in the analytical model to decrease the sound velocity (SI. C). Further, the mechanism of the interaction behind it can be illustrated via a numerical simulation. We have constructed the model in the ***COMSOL MULTIPHYSICS*** © with the FEM approach to study the interaction between sound wave and microstructure (SI. D). As shown in Fig. 5a, the wave is delayed while passing through the inhomogeneous medium that embeds heavy, but "soft", units. The total sound pressure field (Fig. 5a) shows that the wave front is no longer coherent. When the wave enters the highly-dense mesoscopic medium, it interacts with the structure and becomes a diffusive transport. The sound is considerably slowed down compared to the propagation in the pure substrate. Notably, energy is transferred into these units when the sound pressure passes through the structure units. The units have initiated forced vibration. Hence, there is a certain lag. The energy is then transferred back into the substrate and it continues to propagate. Consequently, the propagation of sound has a lag effect. Every structure unit placed in the substrate has an influence on it. Therefore, the filling fraction, size of the units, and structural parameters will be the important ingredients for the sound slowing.

The ratio between the wavelength and the microstructure size of the units set is 100:1, for simulating the non-resonant conditions in our experiment. And the effective parameters such as the Young's modulus and the mass density of both the substrate and the materials in units in the meta-aerogel, is unchanged from the experimental results. The sound pressure in the substrate, in isolation, propagates at a normal speed of about 100 m/s since its components is set as homogeneous porous PAA.

With several manipulations (see SI. D) the transport velocity of the sound at each point along the propagating direction can be attained as shown in Fig. 5. A higher delay of the sound transport velocity is closely related to a larger density ratio (d.r.) of the scattering units to the substrate, with significant contribution, though it reaches its limit (Fig. 5b). Besides, when the modulus ratio (m.r.) between the boundaries of the substrate and the scattering units approaches one, the transport velocity of the sound will be significantly delayed (Fig. 5c).

The interaction process can be demonstrated as the energy carried by the sound wave which is exchanged with the vibration kinetic energy of the attached units (Fig. 5d). This is discussed in the electronic analogy of the analytical model (SI. C).

Fig. 5e clearly shows that the high-level filling fraction ($\alpha$) will enhance the coupling interaction, which contributes to a deeper delay of the sound. Notably, the size of the units (d) plays an extra essential role since the filling fraction will inevitably reach its limit of about 91% in 2D situation. As shown in Fig. 5f, the more coupling interfaces will be enhanced for smaller particle size, thus producing higher interaction degrees.

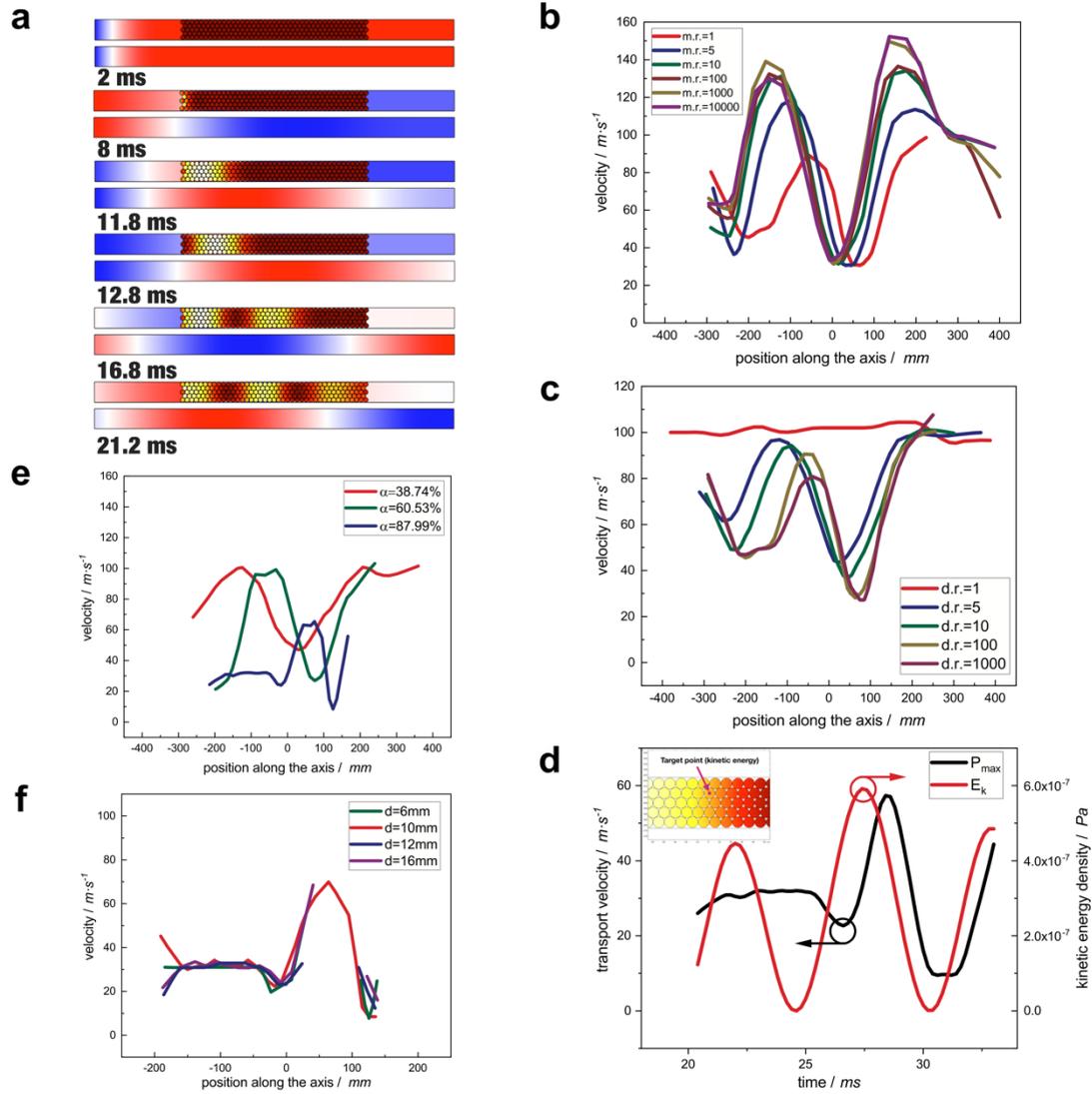

*Figure 5 | Numerical simulation of the sound propagation with strong-coupling interaction. a, The evolution of the propagation in both mediums, and the upper panels shows the analogy of the microstructure of meta-material (heavy units embedding into the soft substrate of porous PAA), and the lower panels display the substrate, whose parameters are the same as those in effective medium mode. b, Influence of mass density ratio (d.r.) on the transport velocity. c, Influence of modulus ratio (m.r.) on the transport velocity. d, Illustration of the energy exchange with the units while*

*transmitting through the structure. **e-f**, Influence of filling fraction (α) and the size of units (d) on the transport velocity of sound.*

The mechanism demonstrates that the method that we have proposed for the slowdown of sound propagation essentially reinforces the non-resonant coupling interaction between the interfaces. Owing to the limits of the filling fraction, the material containing nanoscale microstructures with large amounts of nanoscale soft interfaces will be an agreeable matter to interact with the sound, thus obtaining an ultra-slow sound propagation.

**Conclusion**

We have proposed a novel procedure for preparing a meta-aerogel that has the property of ultra-slow sound. Through an analytical model of effective medium theory, we have found that the most important parameters that influence the sound velocity are the effective modulus of the porous PAA component in our meta-aerogel and the extra dynamic mass density of the scattering units. Furthermore, we have demonstrated that the mechanism of the slow sound is the non-resonant strong coupling interaction between the sound in substrate and the highly concentrated scatter units, whose interface behaves "softly" for an easy transfer of the energy. In basic terms, the slow sound is the result of the non-resonant strong-coupling-interaction-induced deep delay. We achieved the slowest sound velocity in compact solids, at 62 m/s by nano-sizing and adjusting the effective scattering units based on our design.

We have fabricated meta-aerogels with a wide range of mass densities, using the proposed procedure. Their sound velocity is still low, at around 100 m/s, indicating that the meta-aerogels are suitable for designing impedance matching in room acoustic applications. Further, the studies on meta-aerogels can proceed to the next generation of acoustic metamaterials by satisfying the demand for adjustable non-resonant designs of high-contrast sound velocity in composite materials.

**Methods**

*Sol-gel process and pinhole drying method*

The sol-gel technique has been used for the preparation of meta-aerogel, as reported in this paper, with a slightly variant principle. First, 3.096 g $CuCl_2$ is dissolved in the mixed solvent of 30 mL ethanol and 2 mL water to form a precursor solution. 2-3 mL PAA is added to provide the long chain and the dispersion. Here, the addition of the amount of PAA is a crucial parameter related to the target property. Thereafter, the hydrated copper ions $[Cu(H_2O)_6]^{2+}$ is hydrolyzed and poly-condensed to form a copper hydroxide skeleton. The network grows crosslinks, and forms a copper-based alcogel. During the formation of the alcogel, the addition of propylene oxide contributes to the hydrolysis reaction of the copper-based precursors. The aging process will last for 3—4 days at room temperature. Meanwhile, the method of pinhole drying has been used to obtain alcogels of different mass densities. By pinhole drying, the gel of the same raw material ratio can be shrunk to a higher density without structural damage. A meta-aerogel can be obtained after the obtained alcogel is aged, replaced, and subjected to supercritical drying.

*Sound velocity acquisition experiment*

To acquire the sound velocity of the meta-aerogel, we apply the ultrasonic non-destructive testing method. We have non-contact air-coupled ultrasonic probes with the frequency of 400 kHz. First, we collect the sound propagating time without the sample. Then, the sample is placed in the middle of the gap of the probes and the time of sound propagation collected. Thus, a time delay is obtained, which we use to calculate the sound velocity. Further details can be found in the supplementary information (SI. A).

*Numerical simulation with COMSOL MULTIPHISICS$^{©}$*

Our numerical simulation illustrates the sound propagation in the meta-aerogel-like medium. The basic model is composed of a porous substrate and some heavy units embedded in a crystal-like arrangement. The material parameters are set to be similar and adjustable to the material in our real meta-aerogels. The modulus and mass density of the substrate are $6 \times 10^5$ Pa and 62.5 kg/m$^3$, respectively. However, the modulus and mass density of the units are $6 \times 10^5$ Pa and 3500 kg/m$^3$, respectively. The frequency of the incident plane wave is 100 Hz, which implies that the wavelength in the substrate is 1 m, while the diameter of the unit is 10 mm, which is adjustable. Thereafter, we have investigated the whole propagation of sound pressure, by acquiring the transport velocity of the sound pressure in the substrate and the kinetic energy density in the units on the target line along the direction of propagation (Fig. S5c). Further details are given in the supplementary information (SI. D).

**Acknowledgements**

We thank Professor Chen Hong from School of Physics Science and Engineering in Tongji University, Professor Zhang Xixiang from Division of Physical Science and Engineering in King Abdullah University of Science and Technology (KAUST) and Yabin Jin from School of Aerospace Engineering and Applied Mechanics in Tongji University for productive discussions and advices. This work was funded and performed under the financial support from the National Natural Science Foundation of China (No. 11874284) and the National Key Research and Development Program of China (2017YFA0204600).

**Author Contributions**

A.D. and Y.X. contributed equally to this work and should be considered co-authors.
A.D., B.Z. conceived the original idea of constructing the slow sound propagation and preparation of the copper-based meta-aerogels. Y.X. prepared the meta-aerogel samples in this work, conducted the sound velocity acquisition experiments and the numerical simulations. A.D., G.W., B.Z., Y.L. and Y.P. provided the guidance for the consideration of the mechanism. A.D. and Y.X. jointly completed the theoretical derivation.

**Additional information**

Correspondence and requests for materials should be addressed to A.D.

**Competing financial interests**

The authors declare no competing financial interests.

# Ultra-slow sound in non-resonant meta-aerogel

Ai Du[1,†,*], Yuhan Xie[1,†], Yan Li[2], Yongdong Pan[2], Guangming Wu[1], Bin Zhou[1]

## SUPPLEMENTARY INFORMATION

### A. Sound velocity acquisition

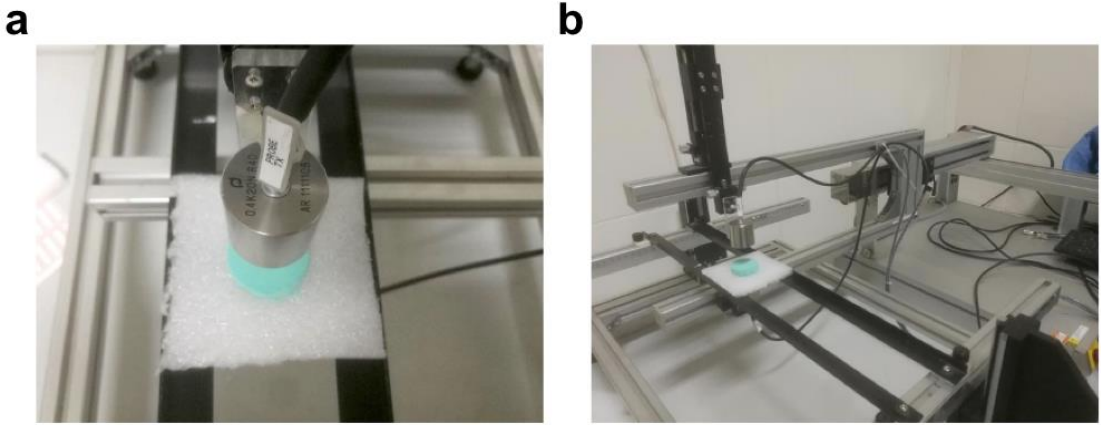

*Figure S1 | Sound velocity acquisition experiment. a, Non-contact air-coupled ultrasonic probe with the frequency of 400kHz. It is worth mentioning that the foam used here has a hold dug in its middle area to let the wave go through. b, Full view of the experiment.*

The device that we have employed to conduct the sound velocity test experiment is shown in Fig. S1. The relation between the time delay and the sound speed in the material is shown in the following as

$$t_1 = \frac{d}{c_0} \tag{1}$$

$$t_2 = \frac{d}{c} \tag{2}$$

$$\Delta t = t_2 - t_1 \tag{3}$$

[1] Shanghai Key Laboratory of Special Artificial Microstructure Materials and Technology, School of Physics Science and Engineering, Tongji University, Shanghai 200092, China
[2] School of Aerospace Engineering and Applied Mechanics, Tongji University, 200092 Shanghai, China
† These authors contributed equally: Ai Du, Yuhan Xie.
* Correspondence and requests for materials should be addressed to Ai Du (email: duai@tongji.edu.cn).

$$c = \frac{d \cdot c_0}{c_0 \cdot \Delta t + d} \qquad (4)$$

where $t_1$ and $t_2$ are the time intervals for sound propagation in the air, with and without samples, respectively. Further, $d$ is the thickness of the sample, $c_0$ the velocity of air-borne sound, and $c$ the sound velocity in the sample.

## B. SAXS analysis of the microstructure of meta-aerogels

### 1. Ways to adjust the microstructure of meta-aerogels

There are multiple ways to adjust the microstructure of meta-aerogels. However, based on our design, the apparent mass density and the PAA-Cu(II) ratio are the parameters that we have mainly applied. The apparent density rises when the structure shrinks without destruction. The change of PAA-Cu(II) ratio behaves as the thickness adjustment of porous PAA wrapping the Cu(II) core, resulting from differential addition of PAA. Furthermore, the filling fraction of structure units changes with the amount of solvent in the state of sol-gel.

### 2. Influence of mass density and the PAA-Cu(II) ratio on the microstructure

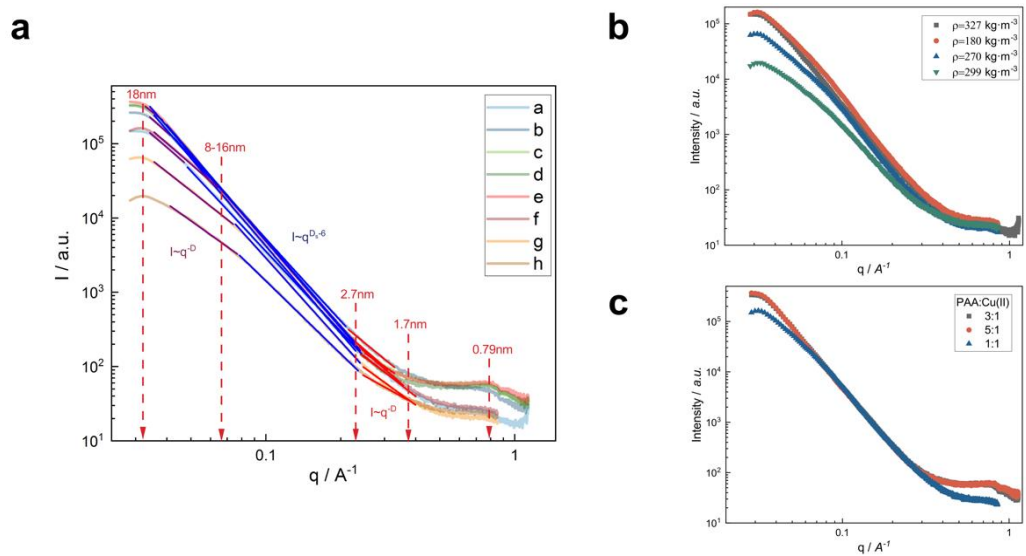

*Figure S2 | Influence of mass density and the PAA-Cu(II) ratio on the microstructure of the meta-aerogels. a, The comprehensive comparison of different meta-aerogels (parameters are shown in the Table S1). b, Influence of mass density on microstructure*

*under the same PAA-Cu(II) ratio. **c**,. Influence of PAA-Cu(II) ratio on microstructure under the similar mass density.*

Prepared from different components ratio and different drying method, the meta-aerogels obey quite similar hierarchical microstructure configuration (Fig. S2a). However, there is still a difference in microstructure crucial to the sound propagation. With the increase of mass density (Fig. S2b), both the surface fractal dimension and the size of the secondary structure becomes smaller, indicating that the secondary structure has been changed into a rougher and denser one, which results in a higher modulus and, subsequently, a higher sound velocity. The microstructure changes (Fig. S2c) along with changes in PAA-Cu(II) ratio. The significant result is the invariance of the surface fractal dimension of the secondary structure that tends to maintain its form. However, the size changes, indicating that the control of adding amount of PAA can properly adjust the thickness of the "soft shell", contributing to the different mass fraction ratio of Cu(II) and PAA in our analytical model.

*Table S1 | Parameters of different meta-aerogel samples.*

| Sample series | PAA-Cu(II) ratio | Mass Density / kg·m$^{-3}$ | Sound Velocity / m·s$^{-1}$ |
|---|---|---|---|
| a | 1:1 | 327 | 178 |
| b | 2:1 | 337 | 293 |
| c | 3:1 | 297 | 305 |
| d | 4:1 | 297 | 353 |
| e | 5:1 | 307 | 333 |
| f | 1:1 | 180 | 65 |
| g | 1:1 | 270 | 109 |
| h | 1:1 | 299 | 129 |

## 3. The SAXS evidence of differences caused by different drying methods

Gibson[1] gives the precise relation between the modulus and mass density of the porous material. However, in meta-aerogels that we have prepared, the microstructure is quite different owing to the drying methods or other probable influences. By employing SAXS method, we further investigate the microstructure of aerogels produced by ambient drying method and the supercritical drying method. We find that the drying methods indeed affect the modulus of the porous material. As discussed in the main text, the meta-aerogel that we have constructed here will have the typical hierarchical microstructure, which can be obtained on the SAXS graphs.

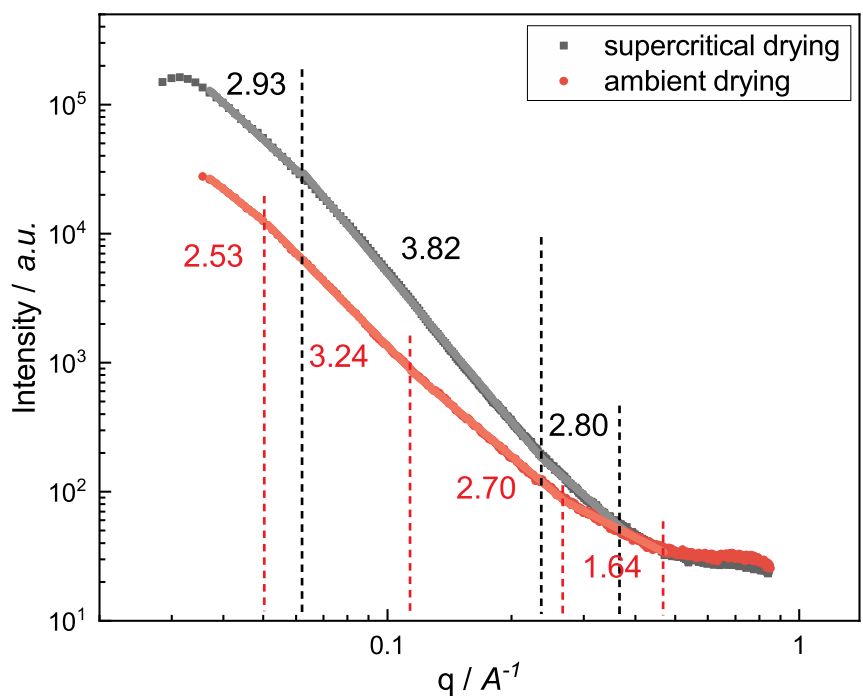

*Figure S3 | Typical SAXS graph of the meta-aerogel produced by different drying methods. The grey and red line are linear fits of different scales for the SAXS data of meta-aerogel produced by the supercritical drying method and ambient drying method, respectively. The numbers labeled on the graph are the fractal dimension in separated scales.*

Although the meta-aerogel is prepared by ambient drying, it still possesses the typical hierarchical microstructure. However, the fractal dimension and the size is changed. Typically, the fractal dimension of the primary structure (shown in Fig. S3) is close to four, implying that the surface of the embedding shell is relatively smooth though it possess large amounts of micro open pores, with a typical size about 8 nm. The SAXS data shows that the fractal dimension is only 3.24 in samples prepared by ambient drying, denoting that the surface is quite rough. Besides, the typical size of the primary particle and primary structure are both larger than those prepared by supercritical drying method. These changes indicate that the "heavy core" grows bigger and the "soft shell" shrinks. Thus, the number of open pores is decreased. The modulus of the porous shell does not necessarily decrease, but the index of its scaling law with the mass density is definitely reduced. Consequently, with reference to the Gibson's theory[1], we denote the index of scaling law relation in ambient-drying meta-aerogels as one for almost closed pores, whereas the parameter is three for open pores in supercritical-drying meta-aerogels.

## C. Analytical model

### 1. Lattice vibration analysis

As is shown in Fig. 4b in the main text, the typical meta-aerogel structure consists of an inner mass-spring cell with the resonant frequency of the oscillators, i.e., $\omega^2 = \frac{k}{m}$, attached to the periodic structure cell. We analyze this chain by considering the motion of the $n^{th}$ unit cell, which is governed by the following equations of motion:

$$(-\omega^2 M + 2K)u_n - K(u_{n-1} + u_n) - k(u_{R_n} - u_n) = 0 \qquad (5)$$

$$(-\omega^2 m + k)u_{R_n} - ku_{R_n} = 0 \qquad (6)$$

where $m$ and $k$ are the mass and spring constant of the oscillator cell, respectively, $u_{R_n}$ is the associated degree of freedom, and $\omega$ the angular frequency of the incident sound wave. By condensing the degree, the equation can read as

$$\left(-\omega^2 M + 2K - \frac{k^2}{k - \omega^2 m} + k\right)u_n - K(u_{n-1} + u_{n+1}) = 0 \qquad (7)$$

The application of Bloch's theorem and a few manipulations lead to the following relation:

$$cosq = 1 - \frac{\omega^2}{2\omega_{PAA}^2}\left(1 + \frac{\frac{m}{M}}{1 - \frac{\omega^2}{\omega_R^2}}\right) \quad (8)$$

where $m/M$ defines the ratio of mass fraction of the attached unit to the primary PAA, and $\omega_{PAA}^2 = \frac{K}{M}$ and $\omega_R^2 = \frac{k}{m}$ represent the Eigen angular frequency of the soft shell PAA and the hard core Cu(II), respectively. Solutions to the above dispersive relation equation for harmonic wave motion, i.e., for assigned values of frequency, provide the wavenumber $q$. Further, the transport velocity $c_t$ can be derived from dividing the angular frequency $\omega$ by the wavenumber $q$ per unit length $l$ as

$$c_t = \frac{\omega l}{q} \quad (9)$$

## 2. Presence of interaction in the analytical model

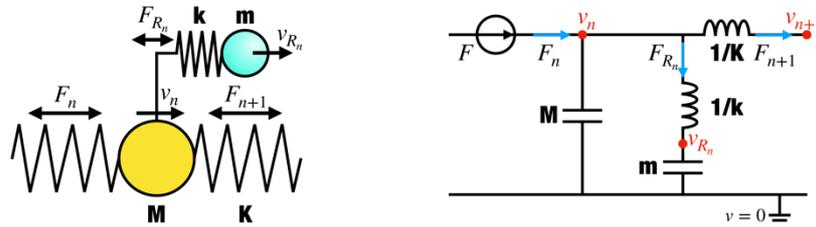

***Figure S4 | Electronic analogy of the analytical model.** By thinking of the force transported on the dynamical chain as the current transported in the circuit, we can get a vivid view of the model that takes into consideration the interaction of the substrate and the attached unit.*

To seize a more intuitive view of the model, we draw an analogy to the concept of electronics, by "replacing" the spring and the mass with an inductor and capacitor, respectively, the stress in the chain is seen as the current while the velocity difference of particles are compared to the electric potential, as

$$M \leftrightarrow C_1, m \leftrightarrow C_2 \quad (10)$$

$$\frac{1}{K} \leftrightarrow L_1, \frac{1}{k} \leftrightarrow L_2 \tag{11}$$

$$F \leftrightarrow I, \tag{12}$$

$$\Delta v \leftrightarrow U \tag{13}$$

Thus, due to the Kirchhoff current law (KCL) and the rules of static AC circuit, there will be

$$I_{in} = I_{out} \tag{14}$$

$$I = j\omega C U \tag{15}$$

$$U = j\omega L I \tag{16}$$

From Equations (10)–(16), we can derivate the dominant equations for stress and the particle velocity as

$$F_{R_n} = \frac{v_n - v_{R_n}}{j\omega \frac{1}{k}} = j\omega m v_{R_n} \tag{17}$$

$$F_n - F_{n+1} = j\omega M v_n + F_{R_n} \tag{18}$$

$$v_n - v_{n+1} = j\omega \frac{1}{K} F_{n+1} \tag{19}$$

where $F$ and $v$ with a subscript $n$ indicates that the analysis is focused on the $n^{th}$ structure unit of the chain. $F_{n+1}$ and $v_{n+1}$ are the force and particle velocity on the $n+1^{th}$ structure unit, and, $F_{Rn}$ and $v_{Rn}$ are the force and particle velocity on the $n^{th}$ attached unit.

Owing to the periodicity in our analytical model, the relation of force and particle velocity should obey

$$F_{n+1} = F_n e^z, v_{n+1} = v_n e^z \tag{20}$$

where the index $z$ is a complex number, $z=a+jb$. The physical meaning of $b$ is hereafter referred to as the propagation constant, like the wavenumber $q$. The dominant equations of the stress and the particle velocity constitute the determinant equations as given below

$$(1 - e^z)F_n - \left(j\omega M + \frac{k}{j\omega} + \frac{k^2}{j\omega(\omega^2 m - k)}\right) v_n = 0 \tag{21}$$

$$-\frac{j\omega}{K} F_n + (e^{-z} - 1)v_n = 0 \tag{22}$$

which gives a limiting condition as

$$\begin{vmatrix} 1-e^{-z} & -\left[j\omega M + \dfrac{k}{j\omega} + \dfrac{k^2}{j\omega(\omega^2 m - k)}\right] \\ -\dfrac{j\omega}{K} & e^{-z} - 1 \end{vmatrix} = 0 \qquad (23)$$

which makes the wavenumber follow the equation

$$e^z + e^{-z} - 2 + \frac{\omega^2}{\omega_{PAA}^2}\left(1 + \frac{\frac{m}{M}}{1 - \frac{\omega^2}{\omega_R^2}}\right) = 0 \qquad (24)$$

which turns into

$$\cosh(z) = 1 - \frac{*}{2} \qquad (25)$$

$$* = \frac{\omega^2}{\omega_{PAA}^2}\left(1 + \frac{\frac{m}{M}}{1 - \frac{\omega^2}{\omega_R^2}}\right) \qquad (26)$$

That is,

$$\cosh(a)\cos(b) + \sinh(a)\sin(b) = 1 - \frac{*}{2} \qquad (27)$$

The right side of the equation is a real value. Hence, the equation should satisfy

$$a = 0, \cos b = 1 - \frac{*}{2} \qquad (28)$$

Thus, we can deduce the function of the transportation velocity by applying Equation (9), as follows:

$$c_t = \frac{\omega l}{b} = \frac{\omega l}{\arccos\left(1 - \frac{*}{2}\right)} \qquad (29)$$

Then, by applying

$$\omega_{PAA} = 2\pi\sqrt{\frac{K}{M}}, \omega_R = 2\pi\sqrt{\frac{k}{m}} \qquad (30)$$

and the modulus $K$ and mass $M$ can be expressed by the relation of diameter $d$ (equal to the unit length $l$ in this model) and density $\rho_{PAA}$:

$$K = \frac{E_{PAA}A}{d} = \frac{E_{PAA}\pi d}{4} \qquad (31)$$

$$M = \rho_{PAA} \frac{4}{3} \pi \left(\frac{d}{2}\right)^3 = \frac{\rho_{PAA} \pi d^3}{6} \tag{32}$$

The, the eigen angular frequency can be calculated from Eqs. (30)–(32):

$$\omega_{PAA} = N \sqrt{\frac{E_{PAA}}{\rho_{PAA} d^2}} \tag{33}$$

$$\omega_R = N \sqrt{\frac{E_R}{\rho_R d_R^2}} \tag{34}$$

where $\rho_R$ is the mass density, $d_R$ the diameter of the attached unit, and $N$ the parameter related to the geometry with $N=1$, $\sqrt{\frac{3}{2}}$, for one-dimension and sphere, respectively. Here, for the coherence of the whole study, the parameters used in the analytical model are identical to the numerical simulation. Thus, we can get the result that $\omega_{PAA}$ is about $1.2 \times 10^{10}$ rad/s and $\omega_R$ is about $5.2 \times 10^9$ rad/s, both of which exceed the central angular frequency (about $2.5 \times 10^6$ rad/s) of the ultrasonic probe that we have used.

Since the $\frac{\omega}{\omega_{PAA}}$ and $\frac{\omega}{\omega_R}$ are both small quantities, and thus $(1 - \frac{*}{2}) \to 1$, we can use the following approximation:

$$arccos\left(1 - \frac{*}{2}\right) = \sqrt{2} \left(\frac{*}{2}\right)^{\frac{1}{2}} = (*)^{\frac{1}{2}} \tag{35}$$

The $m/M$ is considered as the mass fraction ratio of the components, since the meta-aerogel is mainly composed of these two components as

$$D_1 = 1 - D_2 \tag{36}$$

$$\frac{D_2}{D_1} = \frac{1}{D_1} - 1 \tag{37}$$

$$D_1 = \left(1 + \frac{D_2}{D_1}\right)^{-1} \tag{38}$$

where $D_1$ is the mass fraction of PAA, and $D_2$ the mass fraction of Cu(II). Consequently, Eq. (29) can be changed to

$$c_t = N \left(\frac{E_{PAA}}{\rho_{PAA}}\right)^{\frac{1}{2}} \cdot \left(1 + \frac{D_2}{D_1}\right)^{-\frac{1}{2}} \tag{39}$$

which is equivalent to the result of the analytical model described in the main text. This analysis gives us a more meaningful result that the slowness in sound propagation indeed comes from the attached units, whose mass density is quite important in dominating the deep delay of the diffusive sound transport when compared to that of the substrate and modulus of the coupling spring. The inference is also deducible by the use of simulation method to dig into the microscopic mechanism.

### D. Numerical simulation of the sound propagation in meta-aerogels

#### 1. Animated simulation of the slowing down process

To see how the interaction happened between the structure units and the incident sound, the numerical simulation has been done with the finite element method. The structure units are supposed to possess the same parameters as the "heavy cores", but with different sizes. Units of 10 mm in diameter were placed in the substrate to imitate the real situation in a more regular method. The parameters of the substrate have been set to be the same as those of the porous PAA. The modulus of pure PAA is supposed to be identical to the acrylic plastic of modulus $3.3 \times 10^9$ Pa, and the density of pure PAA is 1090 kg/m$^3$, whereas the density of porous PAA is calculated from the meta-aerogel as 62.5 kg/m$^3$. Subsequently, the modulus of the substrate is set as $6 \times 10^5$ Pa. The incident sound propagation is imitated by plane wave of 100 Hz, which results in a much longer wavelength (~ 1 m) compared to the size of the units.

#### 2. Further analysis of the evolution process

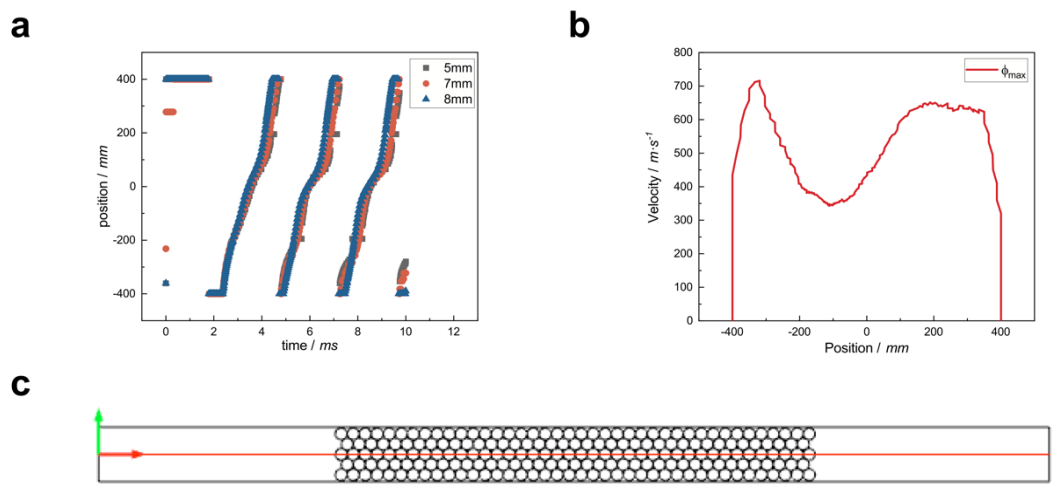

*Figure S5 | The numerical study of the sound propagation. a, Evolution of position of the maximal sound pressure along the propagating direction in the base material. b,*

*Derivate sound transport velocity evolution. **c,** Simulated model and the analysis target line along the propagating direction.*

The steps have been rigorously organized for analyzing the velocity of the simulated sound propagation. First, we acquire the data of the positional evolution of the maximal sound pressure (Fig. S5a) along the axis of the transport direction (Fig. S5c). From the result of the first step we can see that the propagation underwent several periods. Thereafter, we take the derivatives of the positions of the second or later period of evolution time. As shown in Fig. S5b, we have obtained the full view of the velocity at each point of the position and the sound passed along the axis. The position of the maximal sound pressure and the evolution time are in a one-to-one correspondence. Hence, we can finally obtain the figures of sound transport velocity at different positions as shown in the main text.